\documentclass[preprint,showpacs,preprintnumbers,amssymb]{revtex4}

\usepackage{graphicx}
\usepackage{dcolumn}
\usepackage{bm}
\usepackage{cd}
\usepackage{amsfonts}
\usepackage{amsmath}
\usepackage{times}
\usepackage[T1]{fontenc}
\usepackage[applemac]{inputenc}
\usepackage{fancyhdr}

\newcommand{\cdr}{\cd\rightarrow}

\newcommand{\cdd}{\cd\downarrow}

\renewcommand{\d}{{\rm{d}}}
\newcommand{\px}{\partial_{x}}

\newcommand{\pxx}{\partial_{xx}}

\newcommand{\pt}{\partial_{t}}

\newcommand{\pyy}{\partial_{yy}}
\newcommand{\py}{\partial_{y}}
\newcommand{\ptt}{\partial_{t'}}
\newcommand{\ptau}{\partial_{\tau}}

\begin{document}

\preprint{}

\title{Connection between the Burgers equation with an elastic forcing 
term and a stochastic process}

\author{\textbf{E. Moreau and O. Vall\'ee}\\
{\emph{\small{Laboratoire d'Analyse Spectroscopique 
 et d'\'Energétique des Plasmas\\ Faculté des Sciences, 
rue Gaston Berger  BP 4043 \\
18028 Bourges Cedex France}}.}}



\date{\today}

\begin{abstract}
We present a complete analytical resolution of the one dimensional Burgers 
equation with the elastic forcing term $-\kappa^{2} x+f(t)$, 
$\kappa\in\mathbb{R}$. Two methods 
existing for the case $\kappa=0$ are adapted and generalized using variable 
and function transformations, valid for all values of space an time. The 
emergence of a Fokker-Planck equation in the method allows to connect a fluid 
model, depicted by the Burgers equation, with an Ornstein-Uhlenbeck 
process. 
\end{abstract}


\maketitle

\section{Introduction}
Burgers equation is known to have a lot in common with the 
Navier-Stokes equation. In particular it presents the same kind of 
advective nonlinearity, and a Reynolds number may be defined from the 
diffusion term \cite{burg}. In addition, this nonlinear equation is much used 
as model for statistical theories of turbulence from which asymptotical 
behaviours may be determined. But, from an analytical point of view, 
the inhomogeneous form is poor studied, the complete analytic solution being 
closely dependent of the form of the forcing term. For example, the solution of 
the one dimensional Burgers equation with a time-dependent forcing term 
\begin{equation}
    \label{bf1}
    \left| 
    \begin{array}{ll}
    \pt u+u\px u-\nu\,\pxx u=f(t)\\
    u(x,0)=\varphi(x),
    \end{array}
    \right.
\end{equation}
may be obtained by two methods. The first method lies on the  
Orlowsky-Sobczyk transformations (OS) \cite{OS}, where the inhomogeneous 
Burgers equation (\ref{bf1}) is transformed into a homogeneous Burgers 
equation. Nevertheless, there exists an other equivalent method to 
solve analytically this problem. By the way of the well-known Hopf-Cole 
transformation \cite{H-C}, an inhomogeneous Burgers equation may be 
transformed into a linear equation: the heat equation 
with a source term, which is nothing but a Schrödinger equation with an 
imaginary time, and a space and time dependent potential. Then, several 
methods have been developed over past decades to treat this kind of equations. 
One of them, the ``Time-Space Transformation method" (TST), has been 
used in order to solve the Schrödinger equation with a time 
dependent mass moving in a time dependent linear potential (M. Feng 
\cite{TST}). It is thus shown, ref.\cite{ispc}, the equivalence between the 
TST method and the Orlowsky-Sobczyk method, that is to say, the possibility to solve 
analytically by two equivalent ways the Burgers equation with a forcing term in 
$f(t)$. The following diagram resumes this equivalence, where Heat-S designs 
the heat equation with a source term, BE the Burgers equation, and HC 
the Hopf-Cole transformation.

\begin{center}
$\CD
{\rm{Inhomogeneous\ BE}} :f(t) \cdr{ 
{\rm{OS}}}{}{\rm{Homogeneous\ BE}}\\
\cdd{{\rm{HC}}}{}  \cdd{}{{\rm{HC}}}\\
{\rm{Heat-S\,(linear)}} \cdr{{\rm{TST}}}{} {\rm{Heat}}
\endCD$	
\end{center} 
This yields to present this paper as a continuation of the previous 
existing methods. The two latest methods are adapted in order to solve 
the inhomogeneous Burgers equation with a forcing term of the form 
$-\kappa^{2} x+f(t)$, where the  value $\kappa^{2}$ represents the string 
constant of an elastic force. Let us note that Wospakrik and Zen \cite{wos} 
have treated this problem but only in the limiting case where the diffusion 
coefficient tends to zero for the asymptotic mode, whereas the methods 
presented here are valid in all cases. The outline of the paper will be thus 
as follows: the next section is devoted to the treatment of an elastic term, 
firstly by the way of a TST method, and then by using a generalized OS method. 
It is  then shown that a Fokker-Planck equation, associated to the 
Ornstein-Uhlenbeck process, arises in the resolution by the TST method. 
Consequently, an ``adapted'' Hopf-Cole transformation may be 
obtained for this case, which allows physical interpretation in the 
asymptotic limit.

\section{Resolution for an elastic forcing term}

As underlined in the introduction, the TST method allows to solve a 
Schrödinger equation for some kinds of potentials. So the inhomogeneous Burgers 
equation has first to be  transformed into such an equation. Starting 
from the following one dimensional Burgers equation with a linear forcing term 
\begin{equation}
    \label{tet1}
    \left| 
    \begin{array}{ll}
    \pt u+u\px u-\nu\,\pxx u=-\kappa^{2} x +f(t)\\
    u(x,0)=\varphi(x),
    \end{array}
    \right.
\end{equation}
we apply a Hopf-Cole transformation of the form 
$u(x,t)=-2\nu\frac{1}{\Psi(x,t)}\px\Psi(x,t)$ to obtain a heat 
equation with a source term $S$:
\begin{equation}
    \label{tet2}
    \pt\Psi(x,t)=\nu\,\pxx\Psi(x,t)+S(x,t)\Psi,
\end{equation}
where $S(x,t)=\frac{\kappa^{2}}{4\nu}x^{2}-\frac{f(t)}{2\nu}x+c(t)$, $c(t)$ 
being an arbitrary time-dependent function. This kind of equation 
permits to apply a TST method based on several change of variables. In 
\cite{ispc}, and following \cite{TST}, a TST method has been used in order to 
solve a Schrödinger equation with a linear potential. Here, a quadratic 
potential appears in Eq. (\ref{tet2}), so the method will consist this time to put
\begin{equation}
    \label{tet3}
    \Psi(x,t)=P(x,t)e^{h(x,t)},
\end{equation}
with $h(x,t)=a_{1}x^{2}+a_{2}(t)x+a_{3}(t)$ ; $a_{1}$, $a_{2}(t)$ 
and $a_{3}(t)$ being constant or time-dependent functions to be determined.
The transformation (\ref{tet3}) introduced in Eq. (\ref{tet2}) gives
\begin{equation}
    \label{tet4}
    \pt P=\nu\,\pxx P+2\nu\,\px h\,\px P
    +\Big(\nu\,\pxx h+\nu(\px h)^{2}+S-\pt h\Big)P.
\end{equation}
Then, in order to cancel the factor of $P$, we put
\begin{equation}
    \nu\,\pxx h+\nu(\px h)^{2}+S-\pt h=0\ ;
\end{equation}
which gives a polynomial of second degree in $x$. This polynomial becomes
zero since all its coefficients are. It comes respectively
\begin{subequations}
    \label{osgg}
    \begin{gather}
	4\nu a_{1}^{2}+\frac{\kappa^{2}}{4\nu}=0,\label{osgg1}\\
	4\nu a_{1}a_{2}-\frac{f}{2\nu}-\dot{a}_{2}=0,\label{osgg2}\\
	2\nu a_{1}+\nu a_{2}^{2}+c-\dot{a}_{3}=0.\label{osgg3}
    \end{gather}
\end{subequations}
Since Eqs. (\ref{osgg}) are satisfied, Eq. (\ref{tet4}) is simplified 
to
\begin{equation}
    \label{tet5}
    \pt P=\nu\,\pxx P+2\nu\,\px h\,\px P.
\end{equation}
We now apply  to Eq. (\ref{tet5}) the following change of variables
\begin{equation}
    \label{tet6}
    \left| 
    \begin{array}{ll}
    y=r(t)x+q(t),\\
    t'=t.
    \end{array}
    \right.
\end{equation}
This induces a transformation of Eq. (\ref{tet5}) into :
\begin{equation}
    \label{tet7}
    \ptt P=\nu r^{2}\pyy P+
    \Big[(-\dot{r}/r+4\nu a_{1})(y-q)+2\nu r a_{2}-\dot{q}\Big]\py P.
\end{equation}
We have now to cancel the term in $\py P$, so we put
\begin{subequations}
    \label{osggg}
    \begin{gather}
	\dot{r}-4\nu a_{1}r=0\label{osggg1},\\
	2\nu r a_{2}-\dot{q}=0\label{osggg2}.
    \end{gather}
\end{subequations}
Notice that the relation (\ref{osgg1}) gives 
\begin{equation}
    \label{oub1}
    a_{1}={\rm{i}}\frac{\kappa}{4\nu}\,,
\end{equation}
where i$=\sqrt{-1}$, with the result that the solution of Eq. (\ref{osggg1}) 
will be
\begin{equation}
    \label{oub2}
    r(t)=e^{{\rm{i}}\kappa t}.
\end{equation}
Eqs. (\ref{osggg}) being satisfied, we obtain
\begin{equation}
    \label{osg9}
    \ptt P=\nu r^{2}\pyy P\ ;
\end{equation}
and finally the transformation
\begin{equation}
    \label{osg10}
    \tau(t')=\int_{0}^{t'} r^{2}(s)\d s\ ,
\end{equation}
yields to the expected heat equation:
\begin{equation}
    \label{osg11}
    \ptau P(y,\tau)=\nu\,\pyy P(y,\tau).
\end{equation}


We show now that the Orlowsky-Sobczyk method is a particular case of 
the method employed here for an elastic term: the Generalized 
Orlowsky-Sobczyk method (GOS).

\noindent Let us consider again Eq. (\ref{tet1}), and let us introduce 
a new velocity $v\equiv v(x,t)$ such as
\begin{equation}
    \label{osg2}
    u=vr(t)+\alpha x+\psi(t)\ ,
\end{equation}
where $r(t),\alpha,\psi(t)$ are time dependent functions or constant determined 
later. The transformation (\ref{osg2}) introduced in Eq. (\ref{tet1}) yields to :
\begin{equation}
    \label{osg3}
     v\left(\dot{r}+\alpha r\right)+
    x\Big(\kappa^{2}+\alpha^{2}\Big)+
    \Big(\dot{\psi}+\alpha\psi-f\Big)+r\pt v
    +r^{2}v\px v+\alpha r x\px v+r\psi\px v-\nu r\pxx v=0.
\end{equation}
In order to delete the terms in $v$ and $x$, and those only 
depending on time, we put
\begin{subequations}
    \label{osg4}
    \begin{gather}
	\dot{r}+\alpha r=0\label{osgggg1}\\
	\kappa^{2}+\alpha^{2}=0\label{osgggg2}\\
	\dot{\psi}+\alpha\psi-f=0\label{osgggg3}
    \end{gather}
\end{subequations}
Since the system (\ref{osg4}) is verified, then Eq. (\ref{osg3}) is simplified 
into
\begin{equation}
    \label{osg6}
    r\pt v+r^{2}v\px v+\alpha r x\px v+r\psi\px v-\nu r\pxx v=0.
\end{equation}
Then, the same time and space change of variables as Eq. (\ref{tet6}) 
applied to Eq. (\ref{osg6}) leads to
\begin{equation}
    \label{osg8}
    p\ptt v+\left(r\dot{q}+r^{2}\psi\right)\py v+
    (\dot{r}+\alpha r)(y-q)\py v
    +r^{3}v\py v-\nu r^{3}\pyy v=0.
\end{equation}
After what, putting
\begin{equation}
    r\dot{q}+r^{2}\psi=0
\end{equation}
we obtain
\begin{equation}
    \label{osg10}
    \frac{1}{r^{2}}\ptt v+ v\py v=\nu\pyy v\ .
\end{equation}
If we put now $t'$ as
\begin{equation}
    \label{osg11}
    \tau(t')=\int_{0}^{t'}r^{2}(s)\d s\ ,
\end{equation}
it comes a homogeneous Burgers equation governing the new velocity $v$ :
\begin{equation}
    \label{osg12}
    \ptau v+ v\py v=\nu\,\pyy v\ .
\end{equation}
From this, the HC transformation $v=-2\nu\frac{1}{P}\py P$ 
yields again to the expected heat equation
\begin{equation}
    \label{osg13}
    \ptau P(y,\tau)=\nu\,\pyy P(y,\tau)\ .
\end{equation}
Hence, both methods GOS and TST may be connected thanks to a 
commutative diagram like the one of the introduction, with a force 
$-\kappa^{2}x+f(t)$.

\section{Derivation of an Ornstein-Uhlenbeck process}
Let $x(t)$ be a stochastic variable satisfying the following Langevin equation 
and describing an Ornstein-Uhlenbeck process \cite{orn1,orn2}
\begin{equation}
    \label{lang1}
    \frac{\d x}{\d t}=-\kappa x+\sqrt{2\nu}b(t);
\end{equation}
where $b(t)$ stands for a Gaussian white noise verifying the standard 
conditions
\begin{equation}
    \label{lang2}
    \langle b(t)\rangle=0\quad{\rm{and}}\quad
    \langle b(t)b(t')\rangle=\delta(t-t').
\end{equation}
Then, using a Kramers-Moyal expansion, a Fokker-Planck equation may 
be obtained for the transition probability $P(x,t)$ \cite{risken}:
\begin{equation}
    \label{lang3}
    \pt P(x,t)=\kappa\px\left(xP(x,t)\right)+\nu\pxx P(x,t).
\end{equation}
This equation is usually solved by Fourier transform, 
and the solution $P\equiv P(x,x',t)$ for the initial condition 
$P(x,t|x',0)=\delta(x-x')$ reads
\begin{equation}
    \label{lang4}
    P=\sqrt{\frac{\kappa}{2\pi \nu\left(1-e^{-2\kappa t}\right)}}
    \exp\left[-\frac{\kappa\big(x-e^{-\kappa t}x'\big)^{2}}{2\nu
    \big(1-e^{-2\kappa t}\big)}\right].
\end{equation}
It is shown in appendix that this solution may also be found by 
the TST method.\\
The interesting point lies in a connexion existing between the 
Ornstein-Uhlenbeck process (Eq. (\ref{lang3})) and the Burgers equation 
(\ref{tet1}) with $f(t)=0$. In order to see this fact, we apply the 
transformation 
\begin{equation}
    \label{lang8}
    P(x,t)=\Psi(x,t)e^{-\frac{\kappa x^{2}}{4\nu}},
\end{equation}
to the Fokker-Planck equation (\ref{lang3}), which leads to the heat 
equation
\begin{equation}
    \label{lang7}
    \pt \Psi=\nu\pxx \Psi+\left(\frac{\kappa}{2}-
    \frac{\kappa^{2}x^{2}}{4\nu}\right)\Psi.
\end{equation}
So, the Hopf-Cole transformation
\begin{equation}
    \label{lang6}
    u(x,t)=-2\nu\frac{1}{\Psi(x,t)}\px \Psi(x,t),
\end{equation}
transforms Eq. (\ref{lang7}) into the inhomogeneous Burgers equation 
\begin{equation}
    \label{lang5}
    \pt u+u\px u=\nu\pxx u-\kappa^{2} x.
\end{equation}

\noindent This interesting result implies two remarks. Firstly, this connection 
gives rise to a physical meaning of the TST method. Indeed, the function $P$ 
introduced in the transformation (\ref{tet3}) is no more an unspecified 
variable, but takes the sense of a transition probability for the variable 
$x(t)$. Then, considering both Eqs. (\ref{lang8}) and (\ref{lang6}), 
we obtain a relation between the velocity $u$ and the 
transition probability $P$:
\begin{equation}
    \label{rel1}
    u(x,t)=-2\nu\frac{1}{P(x,t)}\px P(x,t)-\kappa x,
\end{equation}
which is composed of a Hopf-Cole part and of a 
linear part. So, this relation may be considered as a 
Hopf-Cole transformation adapted to the Ornstein-Uhlenbeck process.
Moreover, the asymptotic limit of $P(x,x',t)$ is given by 
(\ref{lang4}):
\begin{equation}
    \label{rel2}
    \lim_{t\to\infty}P(x,x',t)=\sqrt{\frac{\kappa}{2\pi\nu}}
    \exp\left(-\frac{\kappa x^{2}}{2\nu}\right),
\end{equation}
and thus, from the relation (\ref{rel1}), we can see that the asymptotic limit 
of the velocity will read
\begin{equation}
    \label{rel3}
    \lim_{t\to\infty}u(x,t)=\kappa x,
\end{equation}
which is a stationary solution. The initial condition 
$P(x,t|x',0)=\delta(x-x')$ expressing the fact that a particle cannot be 
at several positions at the same time, it may be considered as the 
more acceptable condition for $P$. Then, the asymptotic solution 
(\ref{rel3}) have a real physical sense. We can conclude on the fact that an 
elastic forcing term applied to the system gives rise to a stationary transition 
probability in the asymptotic mode. Consequently, the effects of the 
oscillations will decrease, up to disappear in the long time limit, and 
stabilize the system with a velocity proportional to the 
displacement. The evanescence of the effect of the force is due to the 
initial condition sensitivity of the Burgers equation. We can see thereby  on 
the system, a phenomenon closely connected to the turbulence effect: the lost 
of memory in the long-time limit.
\section{Conclusion}
We have presented the complete analytical solution of 
the Burgers equation with an elastic forcing term. The methods 
presented here have been used before but only in the case of a 
time-dependent forcing term. As a perspective, we can say that the 
generalisation of the methods to any order of power of $x$ seems actually be 
a difficult task. Indeed, a transformation of the form $y\to r(t)x+q(t)$, has 
been introduced in order to delete terms proportional to $x$. So this 
transformation seems without effect when higher powers of $x$ appear. 
Moreover, the more the degree will be high, the more the resolution will be 
difficult, due to the increasing number of variables to be introduced. The 
second main result of the paper lies in the existence of links between a 
fluid model (Burgers) and the statistical physics (Ornstein-Uhlenbeck). By a 
set of transformations, we have connected the Burgers equation for the velocity 
$u=\d x/\d t$ to a Fokker-Planck equation for the transition probability of 
the variable $x$. From the Burgers equation (\ref{lang5}), the transformation 
(\ref{rel1}) allows to get directly the Fokker-Planck equation (\ref{lang3}) 
as a specific Hopf-Cole transformation. It appears that the linear force,  
describing the Ornstein-uhlenbeck process, stabilize the system in the 
asymptotic mode with a velocity proportional to the force applied initially, 
since we consider the initial condition $P(x,t|x',0)=\delta(x-x')$ as 
the more acceptable condition. This result shows a characteristic property 
of turbulence, {\it{i.e}} the unpredictability in the long time limit of a 
velocity field governed by the Burgers equation. An application of the methods 
presented here will be described in a forthcoming paper with the case of an 
electric field in a plasma.
\newpage
\appendix*
\section{Solution of the Ornstein-Uhlenbeck process}
\noindent We show that we can recover the solution (\ref{lang4}) by the way of 
our TST method. 

\noindent Rewriting Eq. (\ref{lang3}),  
\begin{equation}
    \label{lang9}
    \pt P=\nu\pxx P+\kappa x\px P+\kappa P,
\end{equation}
we apply the change of variable
\begin{equation}
    \label{lang10}
    \left| 
    \begin{array}{ll}
    y=r(t)x,\\
    t'=t\ .   
    \end{array}
    \right.
\end{equation}
This yields to
\begin{equation}
    \label{lang11}
    \ptt P=\nu r^{2}\pyy P+\left(\kappa -\frac{\dot{r}}{r}\right)y\py P+
    \kappa P.
\end{equation}
To cancel the term in $\py P$ we put obviously
\begin{equation}
    \label{lang12}
    \kappa-\frac{\dot{r}}{r}=0\quad\Leftrightarrow\quad 
    r(t')=e^{\kappa t'}.
\end{equation}
This leads to
\begin{equation}
    \label{lang13}
    \ptt  P=\nu r^{2}\pyy P+\kappa P.
\end{equation}
Then, putting
\begin{equation}
    \label{lang14}
    P(y,t')=\Theta(y,t')e^{\kappa t'},
\end{equation}
followed by the transformation 
\begin{equation}
    \label{lang16}
    \tau(t')=\int_{0}^{t'} r^{2}(s)\d s,
\end{equation}
we obtain the heat equation
\begin{equation}
    \label{chaleur}
    \ptau \Theta=\nu\pyy \Theta.
\end{equation}
Notice that the condition $P(y,y',0)=\delta(y-y')$ implies that
$\Theta(y,y',0)=\delta(y-y')$. The fundamental solution of 
(\ref{chaleur}) is thus 
\begin{equation}
    \label{lang17}
    \Theta(y,\tau)=\frac{1}{\sqrt{4\pi 
    \nu\tau}}\exp\left[-\frac{(y-y')^{2}}{4\nu\tau}\right];
\end{equation}
after what, putting $y$ and $\tau$ in place of their expression, it 
is to say
\begin{equation}
    \label{lang18}
    \left| 
    \begin{array}{ll}
    y=xe^{\kappa t},\\
    \tau=\frac{1}{2\kappa}\left(e^{2\kappa t'}-1\right),
    \end{array}
    \right.
\end{equation}
we obtain 
\begin{equation}
    \label{lang19}
    P=\sqrt{\frac{\kappa}{2\pi \nu\left(1-e^{-2\kappa t}\right)}}
    \exp\left[-\frac{\kappa\big(x-e^{-\kappa t}x'\big)^{2}}{2\nu
    \big(1-e^{-2\kappa t}\big)}\right],
\end{equation}
which is the same result as Eq. (\ref{lang4}).
    
\end{document}